\documentstyle[preprint,aps]{revtex}

\def\bm#1{\mbox{\boldmath$#1$}}
\newcommand{\jtwo}{\mbox{${\cal J}^{(2)}$}}

\tighten
\begin{document}

\title
{COMPETITION BETWEEN T=0 AND T=1 PAIRING IN  PROTON-RICH NUCLEI}

\author{
W. Satu{\l}a$^{1-4}$,
R. Wyss$^{1}$}

\address{
$^1$The Royal Institute of Technology, Physics Department Frescati, 
Frescativ\"agen 24, S--104 05 Stockholm, Sweden \\
$^2$Joint Institute for Heavy Ion Research, Oak Ridge, TN 37831, USA \\
$^3$Department of Physics, University of Tennessee, Knoxville, TN 37996, USA\\
$^4$Institute of Theoretical Physics, Warsaw University,
ul. Ho\.za 69, PL-00-681 Warsaw, Poland 
}

\maketitle
\begin{abstract}
A cranked mean-field model with two-body 
T=1 and T=0 pairing interactions is presented. 
Approximate 
projection onto good particle-number
is enforced via an extended Lipkin-Nogami scheme.
Our calculations suggest the simultaneous presence of both T=0 and T=1 
pairing modes in N=Z nuclei. The  transitions between  different pairing
phases are discussed  as a function of neutron/proton excess, T$_z$, 
and rotational frequency, $\hbar\omega$.
The  additional binding energy due to the T=0 $np$-pairing correlations,
is suggested as a possible microscopic explanation of
the Wigner energy term in even-even nuclei.
\end{abstract}


\narrowtext

The study of pairing correlations is one of a central
theme in nuclear structure physics. Although the energy gain due
to pairing correlations is rather modest, these correlations
strongly influence many  properties of the atomic nucleus.
The large body of phenomena related to pairing among like-particles 
can be well understood, at least qualitatively, in terms of the simple 
BCS model with seniority force.
In light nuclei, especially those with N=Z, 
it is well established empirically that neutron-proton ($np$)
short range (pairing) correlations are of importance. 
The mean-field formalism like
the Hartree-Fock-Bogolyubov (HFB) method is in principle capable
to simultaneously treat both T=0 and T=1 pairing modes. The necessary 
generalizations
of the Bogolyubov transformation were worked out by Goswami and coworkers 
\cite{[Gos64],[Gos65],[Che67]} and Goodman \cite{[Goo72]}.
These early calculations indicated that the importance of
$np$-pairing is restricted to the vicinity of the N=Z line,
\cite{[Wol71]}, and that the T=0 and T=1 pairing phases are exclusive,
see also \cite{[Goo68],[Goo70]}.
The formalism was further extended to describe rotating nuclei~\cite{[Nic78]},
suggesting a possible
phase transition between the two pairing modes
at high spin \cite{[Mul82]}.

The recent progress in nuclear
spectroscopy, related to the event of 
highly efficient detector arrays
and development of radioactive
ion beam facilities, is opening up
new avenues to study the nature of nuclear interactions, in particular,
$np$-pairing correlations at the N=Z line.
Phenomena like possible phase transition between different pairing modes
in rapidly rotating nuclei, the    
influence of $np$-pairing on the position of the proton drip line and 
the stability of drip line nuclei due
to the additional binding energy emerging from $np$-pairing are 
becoming important issues in nuclear structure.

The aim of this paper is to investigate  basic features of
$np$-pairing. We
 present a model applicable for 
 pairing-and-deformation self-consistent
cranking calculations and introduce
 a method to restore approximately
the particle-number symmetry. This concept is an extension 
of the so called Lipkin-Nogami method for the case of a non-separable
proton-neutron system. The method is independent of 
the kind of two-body interaction used in the
calculations. Applying approximate number-projection,
results in the simultaneous presence of both T=0 and T=1 pairing modes.
A detailed discussion of different aspects of our model 
will be given in a subsequent publication.

The starting point of our calculations are the 
eigenstates of a {\it deformed} phenomenological
single-particle potential and therefore, spherical symmetry is broken
already from the beginning.
The main goal is 
to construct a formalism which is flexible enough to account for 
simultaneous scattering of ({\it i\/})
the nucleonic
pairs where both particles
occupy states of different signatures
and
({\it ii\/}) 
pairs where both particles occupy states of the same signature. 
According to Ref.~\cite{[Goo72]}, these two pairing modes 
will be further denoted as
${\bm {\alpha}}\tilde{\bm {\beta}}$ and {\bm {\alpha\beta}},
respectively.

The most general Bogolyubov transformation can be written
as:
\begin{equation}\label{eq3}
\hat\alpha^\dagger_j = \displaystyle\sum_{\alpha >0} 
(U_{\alpha j} a^\dagger_{\alpha}
+V_{\tilde\alpha j} a_{\tilde\alpha}+ U_{\tilde\alpha  j} 
a^\dagger_{\tilde \alpha}
+V_{\alpha j} a_{\alpha}) 
 \end{equation}
where $\alpha (\tilde\alpha)$  denote {\it single particle}
states (including isospin indices) of signature $r=-i(+i)$ respectively, 
while $j$ labels quasiparticles.
Following the calculations of Ref.~\cite{[Nic78]},
we will further impose the so called 
{\it antilinear simplex symmetry}, 
$\hat S_{z}^{A}=\hat P\hat T\hat R_{z}$ 
as a self-consistent symmetry (SCS), see also \cite{[Goo74]}.
One should bear in mind  that, due to the antilinearity of $\hat S_{z}^{A}$,
the transformation properties of creation and destruction operators 
with respect to $\hat S_{z}^{A}$ will depend on the phases of 
the single-particle states.
In other words one cannot introduce any new quantum number 
 associated with that symmetry. 
After applying the $\hat S_{z}^{A}$ symmetry the 
Bogolyubov transformation still remains complex but
the imaginary part
decouples from the real part in the sense  that the different
signature blocks of the
density matrix, ${\bm {\rho=V}}^* {\bm V}^T$, and  pairing tensor
${\bm {\kappa = V}}^*{\bm U}^T$ are either real or imaginary:
\begin{equation}
\label{eq7}
{\bm \rho}= \left( \begin{array}{cc}  
             {\bm \Re}(\rho_{\alpha\beta}) & 0 \\
             0 & {\bm \Re}(\rho_{\tilde\alpha\tilde\beta}) 
      \end{array} \right) +
 i   \left( \begin{array}{cc} 
             0 & {\bm \Im}(\rho_{\alpha\tilde\beta}) \\
             {\bm \Im}(\rho_{\tilde\alpha\beta}) & 0
     \end{array} \right)
\end{equation}
\begin{equation}\label{eq8}
{\bm \kappa}= \left( \begin{array}{cc}  
             0 & {\bm \Re}(\kappa_{\alpha\tilde\beta})  \\
             {\bm \Re}(\kappa_{\tilde\alpha\beta}) & 0 
      \end{array} \right) +
 i   \left( \begin{array}{cc} 
             {\bm \Im}(\kappa_{\alpha\beta}) & 0\\
             0 & {\bm \Im}(\kappa_{\tilde\alpha\tilde\beta})
     \end{array} \right)
\end{equation}
Furthermore, the complex structure  of the single particle potential,
${\bm h}$, and the pairing potential, ${\bm \Delta}$,
and consequently the HFB equations are fully determined by the complex 
structure of the ${\bm \rho}$ and ${\bm \kappa}$ matrices, 
respectively.

In this work we restrict the two-body $np$-pairing interaction to 
a simple extension of the standard seniority pairing
interaction. It is separable in the 
particle-particle channel, ${\bar v}_{\alpha\beta\gamma\delta}\propto
g_{\alpha\beta}g^*_{\gamma\delta}$, with $g_{\alpha\beta}$ proportional
(up to a phase factor) to the overlap 
$\langle\alpha_\tau | \beta_{\tau'}\rangle$ between 
single-particle wave functions.  
Apart from weak modifications due to the isovector components
of the nuclear-mean field, like the static Coulomb potential, 
the interaction is dominated by 
${\bm {\alpha}}\tilde{\bm {\alpha}}$  and {\bm {\alpha\alpha}} 
types of
pairing. 

In order to relate  the structure of the
$\bm{\alpha\tilde\alpha}$ and $\bm{\alpha\alpha}$ pairing modes to
the isospin quantum numbers, let us briefly consider
a nucleus with isospin- and time-reversal symmetry.
By decomposing the pairing potential
into the different isospin 
components (T,T$_z$) one finds 
that the T=1 and
T=0 components of the $np$-pairing  depend on the combinations of
the same elements of the pairing tensor but with opposite sign~\cite{[Goo72]}.
Consequently, with the pairing tensor of the form of  (\ref{eq8}), 
the T=0 
component of ${\bm {\alpha}}\tilde{\bm {\alpha}}$ pairing
is ruled out due to the $\hat S_{z}^{A}$
symmetry. Similar analysis shows that T=1 component of the 
${\bm \alpha}{\bm \alpha}$ pairing also vanishes. 
Therefore, in our model the $\bm{\alpha\tilde\alpha}$ pairing is
equivalent to T=1 and $\bm{\alpha\alpha}$ to T=0.
This simple analysis also reveals the {\it important\/} role played
by {\it symmetries\/} 
in the theoretical description of $np$-pairing, see also \cite{[Wol71]}.

The average gap parameters are equal to
(we adopt the 
convention where $\tau= 1$ for neutrons and $\tau=-1$ for protons)
\begin{equation}
\Delta^{T=1}_{\alpha_\tau ,\widetilde{\beta_\tau}} =
-\delta_{\alpha_\tau \beta_\tau} \Delta^{T=1}_{\tau\tau}\quad \mbox
{where} \quad
\Delta^{T=1}_{\tau\tau}=G_{\tau\tau}^{T=1}\displaystyle\sum_{\alpha_\tau > 0}
\kappa_{\alpha_\tau ,\widetilde{\alpha_\tau}}, 
\end{equation}
for T=1 $pp$- ($nn$-) pairing,
\begin{equation}\label{del1}
\Delta^{T=1}_{\alpha_\tau ,\widetilde{\beta_{ -\tau}}} = 
        - \langle \alpha_\tau | \beta_{ -\tau}\rangle 
            \Delta_{np}^{T=1}   
\quad \mbox{where} \quad \Delta_{np}^{T=1} =
{1\over 2}G^{T=1}_{np} \displaystyle\sum_{\alpha_n ,\beta_p > 0}
\langle \alpha_n |\beta_p\rangle \left\{ 
\kappa_{\alpha_n ,\widetilde{\beta_p}} 
+ \kappa_{\beta_p ,\widetilde{\alpha_n}}, \right\}  
\end{equation}
for T=1 (${\bm \alpha}\tilde{\bm\alpha}$) $np$-pairing and
\begin{equation}\label{del0}\begin{array}{ccc}
&\Delta^{T=0}_{\alpha_\tau ,\beta_{ -\tau}} = 
          i \tau \langle \alpha_\tau | \beta_{ -\tau}\rangle 
            {\bm \Im} ( \Delta_{np}^{T=0})   \quad \mbox{and}
\quad
\Delta^{T=0}_{\widetilde{\alpha_\tau} ,\widetilde{\beta_{ -\tau}}} = 
          -i \tau \langle \alpha_\tau | \beta_{ -\tau}\rangle 
            {\bm \Im} ( \Delta_{np}^{T=0})  & \\
&\mbox{where}\quad \Delta_{np}^{T=0}   
={i\over 2} G^{T=0}_{np} \displaystyle\sum_{\alpha_n ,\beta_p > 0}
\langle \alpha_n |\beta_p\rangle \left\{
{\bm \Im} (\kappa_{\widetilde{\alpha_n} ,\widetilde{\beta_p}})
- {\bm \Im} (\kappa_{{\alpha_n} ,{\beta_p}} ) \right\}  &
\end{array}\end{equation}
for T=0 (${\bm \alpha}{\bm\alpha}$) $np$-pairing.
The strengths of the interaction
are denoted by $G^T_{\tau\tau '}$.

To prevent a sudden collapse of the static pairing correlations,
e.g. induced by fast 
nuclear rotation
we introduce an approximate particle-number projection using the 
Lipkin-Nogami (LN) method~\cite{[Pra73]}.
This method is equivalent to a restricted HFB-type variation, 
$\delta \langle HFB | \hat {\cal H}^\omega | HFB \rangle =0$,
for the Routhian:
\begin{equation}\label{eq13}
\hat {\cal H}^\omega = \hat H^\omega  - \displaystyle\sum_{\tau} 
\lambda^{(1)}_\tau
\Delta\hat N_\tau - \displaystyle\sum_{\tau \tau'} \lambda^{(2)}_{\tau\tau'}
\Delta\hat N_\tau \Delta\hat N_{\tau'}. 
\end{equation}
In the LN method  the parameters $\lambda^{(1)}_\tau$ 
are standard Lagrange-type multipliers
  whereas the parameters 
$\lambda^{(2)}_{\tau\tau'}$ are kept constant during the variational 
procedure and eventually adjusted self-consistently using three 
additional subsidiary conditions.
\begin{equation}\label{eq14} 
\langle \hat{\cal H}^\omega(\Delta\hat N_\tau\Delta\hat N_{\tau'} - \langle
\Delta\hat N_\tau \Delta\hat N_{\tau'} \rangle )\rangle = 0.
\end{equation}
where $\Delta\hat N_{\tau}\equiv \hat N_{\tau} - N_\tau$
and the symbol $\langle\quad\rangle$ stands for the average over
the $|HFB\rangle$ state.
The LN theory is technically similar to the HFB
theory but for the Routhian (\ref{eq13}). The resulting 
LN equations take the form of HFB equations with a single-particle 
field and pairing field renormalized as follows:
\begin{equation}\label{eq21}
h_{\tau\tau'}^{LN} \rightarrow h_{\tau\tau'} + 2\lambda^{(2)}_{\tau\tau'}
\rho_{\tau\tau'} \quad \mbox{and} \quad 
\Delta_{\tau\tau'}^{LN} \rightarrow \Delta_{\tau\tau'} - 
2\lambda^{(2)}_{\tau\tau'} \kappa_{\tau\tau'}. 
\end{equation}

An open question in mean-field calculations with $np$-pairing is related
to the strength of the interaction. Whereas 
the strength of the $pp$- and $nn$- seniority  pairing force is well 
established by a fit
to the odd-even mass differences, very little is known about the strength 
of the
$np$-pairing force. Based on isospin-symmetry arguments, 
it seems
well justified to assume 
that at the N$\sim$Z line $G^{T=1}_{pp(nn)}\sim G^{T=1}_{np}$.
Therefore, our results will be presented either as a function of or at a 
given
value of the parameter $x^{T=0}$ that
scales the strength of T=0 $np$-pairing with respect to the average strength
calculated for $nn$- and $pp$- pairing correlations i.e. 
$x^{T=0}=G_{np}^{T=0}/G_{np}^{T=1}$
while $G_{np}^{T=1}=(G^{T=1}_{nn}+G^{T=1}_{pp})/2$.

Fig.~1 shows the pairing gaps at zero frequency for a self-conjugate, N=Z,
nucleus calculated with (BCSLN) and without (BCS) 
approximate number projection
as a function of $x^{T=0}$. In this case  we also disregard
the Coulomb interaction and, therefore,
$\Delta_{pp}^{T=1}=\Delta_{nn}^{T=1}=\Delta_0$
and $G_{pp}^{T=1}=G_{nn}^{T=1}=G_{np}^{T=1}$.
The BCS version of our model has been
discussed in the literature \cite{[Goo72]} and
the solutions can be characterized as  follows:
({\it i\/}) For $x^{T=0}<1$ ($G^{T=0}_{np} < G^{T=1}_{np}$) the T=1 pairing 
is energetically  favoured over the T=0 pairing. The pairing energy 
depends only 
on  $\Delta^2\equiv 2\Delta_0^2 + (\Delta_{np}^{T=1})^2$
and {\it no} energy is gained by activating the T=1 $np$-pairing.
({\it ii\/}) The solution at   $x^{T=0}=1$ ($G^{T=0}_{np}=G^{T=1}_{np}$) is 
highly degenerate. The HFB energy depends only on 
$\Delta^2\equiv 2\Delta_0^2 + (\Delta_{np}^{T=1})^2
+ |\Delta_{np}^{T=0}|^2$. Also in this limit, no 
energy is gained due to $np$-pairing.
({\it iii\/}) The solution at $x^{T=0}>1$
($G^{T=0}_{np} > G^{T=1}_{np}$)
corresponds to a pure T=0 $np$-pairing phase.

As shown in Fig.~1, the results from the number-projected calculations
are quite different.
The critical value of the strength necessary to activate T=0 $np$-pairing 
is larger, $x^{T=0}_{crit}\approx 1.1$. 
The LN-method introduces different modifications of
 the pairing potential 
for the T=1 $pp$- ($nn$-) and
 T=1 $np$-pairing field. It causes that at
$x^{T=0} < x^{T=0}_{crit}$  and
under the assumption of $G_{np}^{T=1}=G_{\tau\tau}^{T=1}$ the 
pairing gap for the T=1 $np$-pairing, $\Delta_{np}^{T=1}$, becomes zero. 
It implies that the LN method has
an isovector component, that requires further investigation.
However, at  $x^{T=0} > x^{T=0}_{crit}$  the  T=0 $np$-pairing correlations 
{\it coexists\/} with the T=1, $|$T$_z|$=1  
pairing. The exclusivness of T=0 and T=1 pairing phases in N=Z nuclei
is a generic feature of the BCS method and is smeared out in the 
number-projected calculations.

For N$\ne$Z both BCS and BCSLN models provide  solutions 
which are qualitatively similar to the T$_z$=0, BCSLN case
i.e. the T=0 $np$-pairing correlations coexist with the T=1 $pp$-
and $nn$-pairing correlations.  
It is worth stressing that
the value of $x^{T=0}_{crit}$ is strongly T$_z$ dependent
i.e. increases quite rapidly with neutron/proton excess or, alternatively,
$np$-correlations are restricted to small $|$T$_z|$. 
In consequence, our calculations suggest a
critical value of  $|$T$_z^{crit}|$ beyond which, i.e., for 
$|$T$_z|>|$T$_z^{crit}|$, there is no collective solution to
$np$-pairing, see Fig.~2.

The additional binding arising from T=0, $np$-correlations 
as well as the narrow region where it is active, is shown
in Fig.~2. Early calculations of nuclear masses
based on a macroscopic-microscopic approach
have shown particularly strong deviations from the experimental
data in the vicinity of the N$\sim$Z line~\cite{[Mye66]}.
Most probably, only part of 
these deviations can be attributed to $np$-pairing while part
of it can be accounted for by the self-consistent mean-field. 
One can assume that the single-particle mean field is
properly taken
into account by the extended Thomas-Fermi model
which is a semi-classical approximation to the Hartree-Fock 
method~\cite{[ETFSI]}.
In this model there is a systematic binding energy offset by
$\sim 2$\,MeV
at  the N=Z line that one might associate
to the lack of $np$-pairing correlations.
Based  on this assumption and the results of \cite{[ETFSI]}  we can estimate 
the strength of the T=0 pairing force
$G^{T=0} \approx 1.2 G^{T=1}$ as shown in Fig.~2. The shell-model
estimate of Ref.~\cite{[Ros48]} yields
$G^{T=0} \approx 1.3 G^{T=1}$.
Modern versions of macroscopic-microscopic mass calculations~\cite{[Mol95]}  
cure the difficulties arising around the N$\sim$Z line 
by introducing an extra term to the liquid drop formula - 
the so called Wigner energy~\cite{[Wig37]}. 
Hence, a possible microscopic origin of the Wigner term in even-even 
nuclei are
the T=0 $np$-pairing correlations.

A third way to generate a phase transition from
T=1 to T=0 pairing is by rotation~\cite{[Nic78],[Mul82]}.  The results 
of cranking calculation 
for   $^{46,48}$Cr are shown in
Fig.~3. These are qualitative
calculations at constant deformation and as such should not directly be 
compared to 
the experimental data.
The  calculations are performed with and without the 
T=0 force. At $\hbar\omega=0$,
the strength of the T=0 force is undercritical and T=0 pairing is not active.
At a certain critical frequency, $\hbar\omega_{crit}$, there is a sudden
$\it onset\/$ of T=0 pairing, see also~\cite{[Mul82]}.
The latter effect can be viewed as either a phase transition or
band crossing.  The coherent action of centrifugal and Coriolis
forces tend to align the angular momentum of 
the quasi-particles along the rotational axis.
It weakens the T=1 pairing correlations and simultaneously,
increases the number of pairs of 
nucleons with parallel coupled
angular momenta, thus enforcing the T=0, ${\bm \alpha}{\bm \alpha}$,  
$np$-pairing correlations.  
In the T=0 phase, angular momentum is built by smoothly
aligning $np$-pairs               
along the rotational axis, without involving any pair breaking mechanism. 
This situation 
is totally different from the well known response of the
T=1 pairing to nuclear rotation, where pairs are broken
to generate angular momentum.
With increasing frequency,
the T=0 pairing correlations tends to saturate.
Note that
the T=0 phase coexists with T=1 $pp$- and $nn$-pairing phases although
the onset of T=0 $np$-correlations suppresses the T=1 phase.
However, our calculations show  that the T=0 $np$-pairing and T=1 
$np$-pairing 
phases are always exclusive at $\hbar\omega\ne0$, 
independently on their relative strengths.

The influence of the T=0 $np$-pairing correlations
on the dynamical moment of inertia is conspicuous. Nuclear rotation 
in the presence of T=0 pairing correlations resembles  classical 
rigid body like rotation even 
though T=1 $pp$- and $nn$-pairing correlations are present.  Note also that  
the moments of inertia  at large frequency 
in the presence of T=0 pairing exceed by far the value
obtained for a system without pairing, see Fig.~3. 
Even though the situations discussed above were visualized for $^{46,48}$Cr
only, these classes of solutions appear to be generic for all even-even 
nuclei with N$\sim$Z, i.e. do not depend qualitatively on A.

Our results can be summarized as follow:
The previously suggested exclusiveness of the T=0 and T=1 pairing
phases \cite{[Goo68],[Goo70],[Mul82]} 
does not find support in our calculations.
The sudden phase transition between the T=0 and T=1 pairing modes is
a generic feature of the BCS approximation for N=Z nuclei.
This phase transition becomes smeared out in number-projected
LN calculations. There, the T=0 $np$-pairing correlations coexist
with T=1 $nn$- and $pp$-pairing correlations over a
broad range
of the strength $G_{np}^{T=0}$  as well as
rotational frequency $\hbar\omega$, when $x^{T=0}>x^{T=0}_{crit}$. 
However, 
pairing correlations of 
  ${\bm \alpha}{\bm \alpha}$
  and ${\bm \alpha}\tilde{\bm \alpha}$ type counteract.
For N$\ne$Z, the T=0 $np$-pairing correlations and T=1 
$nn$- and $pp$-pairing correlations do coexist in both BCS and 
number-projected LN calculations.
The T=0 $np$-pairing correlations are confined to a
narrow region along the N=Z line. 
The additional binding arising from these correlations may 
be viewed as a microscopic
origin of the Wigner term in even-even nuclei.
  Even in the cases where $np$-pairing correlations are not present in
  the ground state one can
  generate T=0 $np$-pairing correlations 
  at large rotational frequencies.
  An onset of ${\bm \alpha}{\bm \alpha}$ $np$-pairing quite dramatically
  influences the mechanism of building angular momenta. The T=0
  $np$-pairs can easily be  decoupled from the deformed
  core by the Coriolis force and consequently nuclear rotation resembles 
  rigid-body like rotation though $pp$- and $nn$-pairing 
correlations are present.
In both BCS and number-projected calculations, the
T=0 and T=1 phases of $np$-pairing are exclusive at $\hbar\omega\ne0$.
We believe that this 
  exclusiveness is an artifact related to our schematic interaction 
and/or the assumed self-consistent symmetries.

 ---------------------------

We wish to thank  J. Dobaczewski, A. Goodman and  W. Nazarewicz for many
stimulating discussions. 
The Joint Institute for Heavy Ion Research has as member institutions
the University of Tennessee, Vanderbilt University, and the Oak Ridge 
National Laboratory. This research was supported in part by the U.S.
Department of Energy (DOE) through Contract No. DE-FG05-93ER40770 with 
the University of Tennessee, by the 
G\"oran Gustafsson Foundation, by the Swedish Natural Science Research Council
(NFR) and by the Polish State Committee for Scientific Research (KBN) under 
Contract No. 2 P03B  034 08.
One of us (W.S.) would like to express his sincere thanks  
to the DOS's Institute for Nuclear Theory in Seattle for financial support 
during INT'95 workshop where part of this study was completed.

\vspace{1cm}

{\bf Figure captions}

\vspace{1cm}

{\bf Fig.~1}  Average pairing gaps for a self-conjugate nucleus,
N=Z, as a function of 
$x^{T=0}=G_{np}^{T=0}/G_{np}^{T=1}$.
Left (right) panel shows the results of calculations 
without (with) particle number projection.

{\bf Fig.~2}  Calculated additional binding energy,
$E(x^{T=0})-E(x^{T=0}=1)$, arising from the
presence of T=0 pairing for a number of Cr-isotopes in the vicinity
of the T$_z$=0 line. Different curves denote results
for: $x^{T=0}$=1.1~($\bullet$),
1.2~($\triangle$), 1.3~(solid triangles) and 1.4~($\star$). 
The solid line without symbols denotes the Wigner 
energy term due to \cite{[Mye66]} and the dotted line ($\ast$)
marks the result of the ETFSI-model~\cite{[ETFSI]}.

{\bf Fig.~3}  The calculated dynamical moments of inertia for 
$^{46}$Cr and $^{48}$Cr. Open circles correspond to the case of pure 
T=1, T$_z$=$\pm$1 pairing and the sharp peak in
\jtwo~ 
is due to the breaking of the $f_{7/2}$ pairs. 
The curve marked by solid dots indicate calculations with undercritical
T=0 $np$-pairing strength at $\hbar\omega$=0.
The sudden rise of the moment of inertia
corresponds to the critical frequency, where the T=0 pairing correlations
switch on.
Note the entirely different behaviour of the
moment of inertia for the two cases. The dotted line ($\ast$) 
denotes the calculations without pairing.


\begin{thebibliography}{10}

\bibitem{[Gos64]}
{A. Goswami, Nucl. Phys. {\bf 60} (1964) 228}.

\bibitem{[Gos65]}
{A. Goswami and L.S. Kisslinger, Phys. Rev. {\bf 140 } (1965) B26}.

\bibitem{[Che67]}
{H.T. Chen and A. Goswami, Phys. Lett. {\bf B24} (1967) 257}.

\bibitem{[Goo72]}
{A.L. Goodman, Nucl. Phys. {\bf A186} (1972) 475}.

\bibitem{[Wol71]}
{H. Wolter, A. Faessler and P. Sauer, Nucl. Phys. {\bf A167} (1971) 108}.

\bibitem{[Goo68]}
{A.L. Goodman, G.L Struble and A. Goswami, Phys. Lett. {\bf B}26 (1968) 260}.

\bibitem{[Goo70]}
{A.L. Goodman, G.L Struble, J. Bar-Touv and A. Goswami, Phys. Rev. {\bf C}2
  (1970) 380}.

\bibitem{[Nic78]}
{K. Nichols and R.A. Sorensen, Nucl. Phys. {\bf A309} (1978) 45}.

\bibitem{[Mul82]}
{E.M. M{\"u}ller, K. M{\"u}hlhans, K. Neerg{\aa}rd and U. Mosel, Nucl. Phys.
  {\bf A383} (1982) 233}.

\bibitem{[Goo74]}
{A.L. Goodman, Nucl. Phys. {\bf A230} (1974) 466}.

\bibitem{[Pra73]}
{H.C. Pradhan, Y. Nogami and J. Law, Nucl. Phys. {\bf A201} (1973) 357}.

\bibitem{[Mye66]}
{W. Myers and W. Swiatecki, Nucl. Phys. {\bf 81} (1966) 1}.

\bibitem{[ETFSI]}
{Y. Aboussir, J.M. Pearson, A.K. Dutta and F. Tondeur, Atomic Data and Nuclear
  Data Tables {\bf 61}, (1995) 127}.

\bibitem{[Ros48]}
{L. Rosenfeld, Nuclear forces, North-Holland, Amsterdam (1948)}.

\bibitem{[Mol95]}
{P. M\"oller, J.R. Nix, W.D. Myers, W.J. Swiatecki, At. Data Nucl. Data Tables
  {\bf 59} (1995) 185}.

\bibitem{[Wig37]}
{E. Wigner, Phys. Rev. {\bf 51} (1937) 106}.

\end{thebibliography}
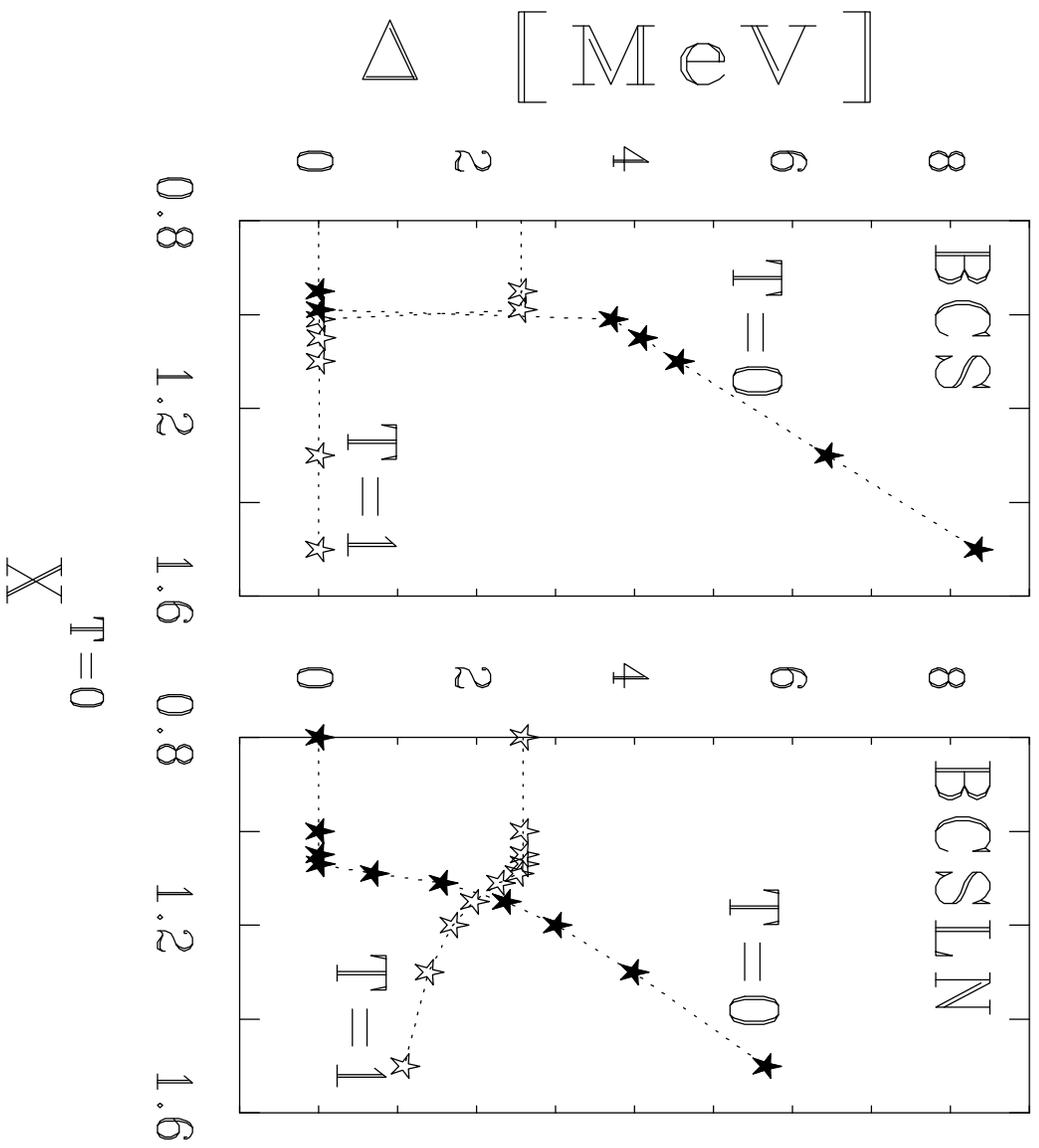
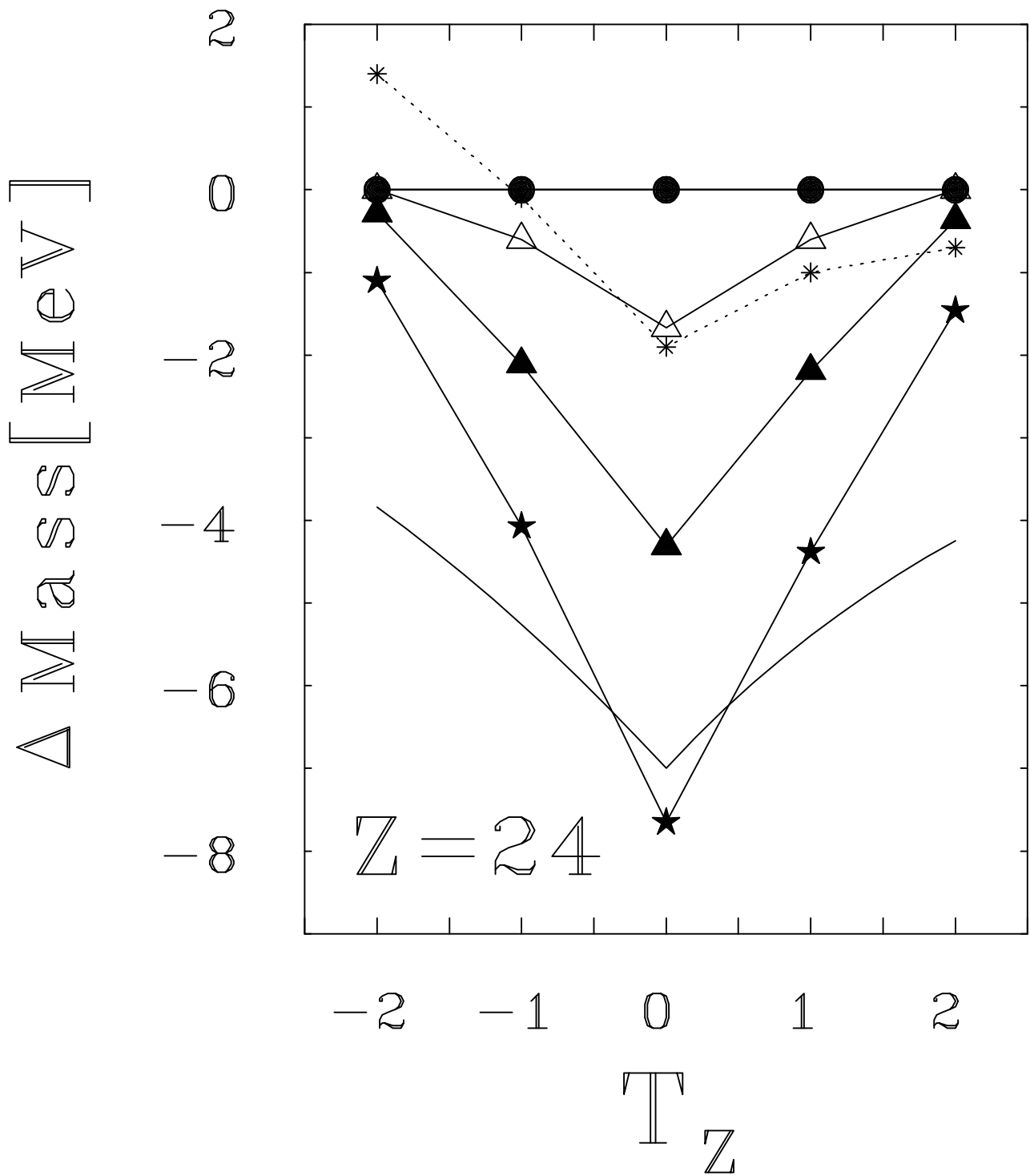
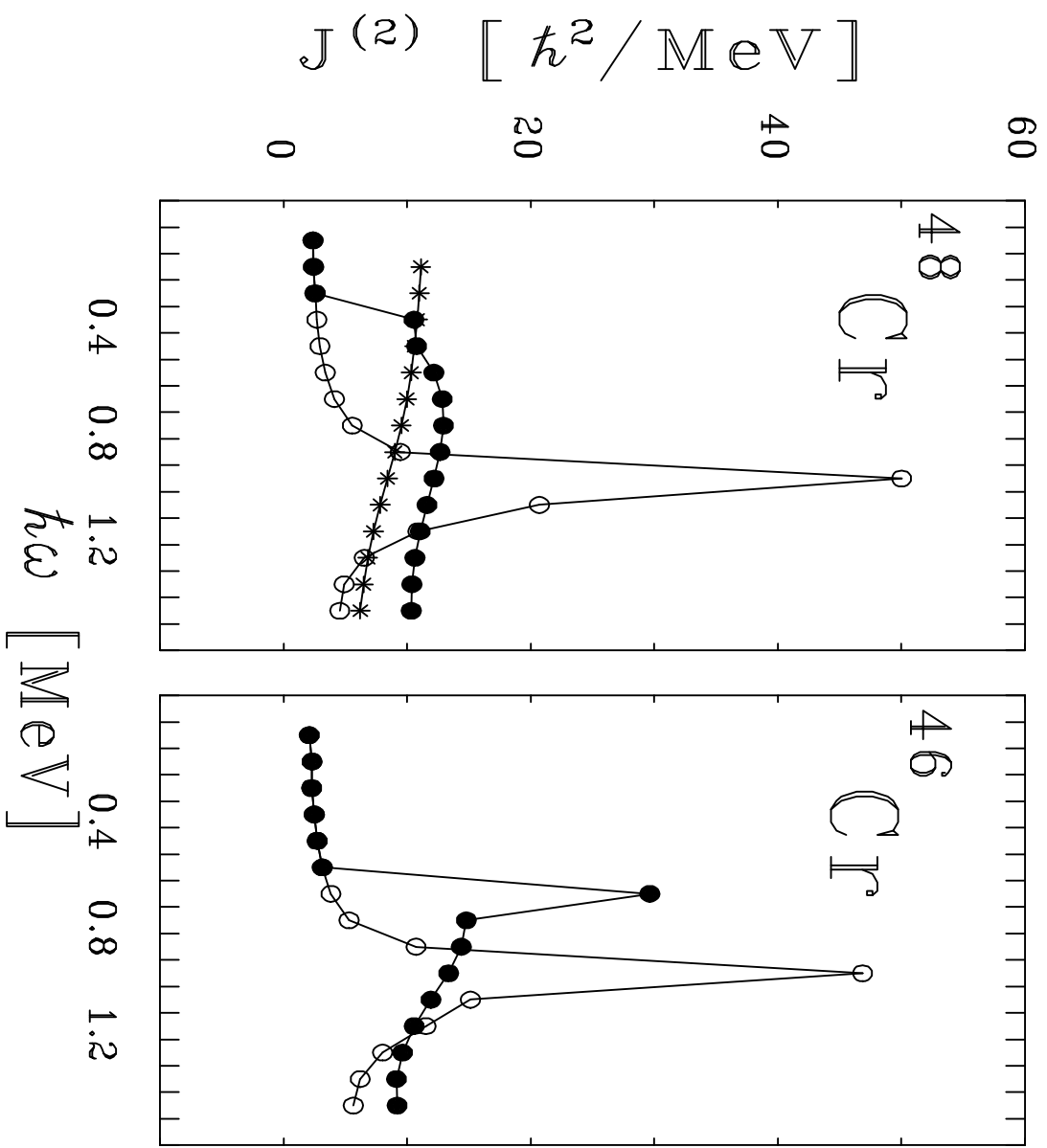
\end{document}